\documentclass[prl,twocolumn,showpacs]{revtex4}
\usepackage{amssymb}
\usepackage{epstopdf}
\usepackage{graphicx}
\usepackage{dcolumn}
\usepackage{amsmath}
\usepackage{amsfonts}

\begin{document}

\title[Short Title]{Selective enhancement of coherent optical phonons using THz-rate pulse train}
\author{Muneaki HASE}
\author{Tetsuya ITANO}
\author{Kohji MIZOGUCHI}
\author{Shin-ichi NAKASHIMA}
\affiliation{Department of Applied Physics, Osaka University, 2-1 Yamadaoka, Suita 565-0871, Japan
}

%\date{\today}

\begin{abstract}
Selective enhancement of coherent optical phonons in Bi-Sb mixed crystals has been performed using terahertz (THz) rate pulse trains generated by spatial phase modulation of femtosecond pulse with a liquid-crystal spatial light modulator. The profile of the shaped pulse train has a flat-toped shape. The Bi-Bi, Bi-Sb, and Sb-Sb vibrations can be enhanced and canceled selectively for pulse trains with a suitable repetition rate. 
\end{abstract}

\pacs{keywords: femtosecond, pulse shaping, coherent phonon, mixed crystal}

\maketitle 
In recent years, coherent phonons have been studied in various materials by using femtosecond pump-probe techniques.\cite{rf1,rf2,rf3,rf4,rf5} The coherent phonons can be generated by ultra short laser pulses with a high degree of temporal coherence. The nature of the coherent phonons has been studied in semimetals\cite{rf1,rf2} and semiconductors.\cite{rf3} Because the coherent phonons are lattice vibrations having the same phase, the control of the amplitude of coherent phonons by using multiple pulses is possible. One of the aims of controlling the coherent lattice vibrations is to promote structural changes that do not occur under ordinary conditions.

Some experimental results for control of the coherent oscillations have been reported. Weiner {\it et al.} have demonstrated that molecular motion generated through impulsive stimulated Raman scattering is enhanced by means of timed sequence of femtosecond pulses.\cite{rf6,rf7} In their work, THz-rate pulse trains have been used for repetitive excitation of a single phonon mode in $\alpha$-perylene molecular crystals. Dekorsy {\it et al.} have demonstrated that the amplitude of the coherent optical phonons can be changed by using double-pulse excitation in GaAs.\cite{rf8} A similar result was obtained in bismuth by Hase {\it et al.} \cite{rf9} The multiple excitation techniques have been applied to control the electro-magnetic radiation emitted from the surface of semiconductors.\cite{rf10,rf11} The results of the multiple excitation of the coherent phonons suggest that the selective enhancement of a particular phonon among a number of phonon modes is possible. 

In this paper we demonstrate both selective enhancement and cancellation of the coherent optical phonons in Bi-Sb mixed crystal systems using THz-rate pulse trains with excitation energies of $\leq$1 nJ/pulse. The Bi$_{1-x}$Sb$_{x}$ mixed crystal system is a three mode type alloy and the $A_{1g}$ and $E_{g}$ type vibrations associated with Bi-Bi, Bi-Sb, Sb-Sb sublattices have been observed.\cite{rf12,rf13} In our pump-probe experiments, three coherent $A_{1g}$ phonons were observed, however $E_{g}$ phonons were not detected. We could selectively enhance and cancel the coherent $A_{1g}$ phonon of Bi-Bi, Bi-Sb, and Sb-Sb vibrations. These experimental results were explained by the fact that the coherent phonons in the mixed crystals are linear combination of damped harmonic oscillations. Furthermore, the ratio of the coherent phonon amplitude to the number of pulse sequence was systematically studied.

The samples used in this study were Bi$_{0.31}$Sb$_{0.69}$ mixed crystals prepared by evaporation on polished silicon substrates at room temperature. The crystal was about 100 nm thick. The femtosecond time-resolved measurements using a reflection type pump-probe set-up were carried out at room temperature. The light source was a mode locked Ti:sapphire laser with a wavelength of 800 nm, providing 60 fs pulses. The pump and probe beams were polarized orthogonal each other to avoid the scattered pump beam. The pump and probe beams were focused on the samples to a diameter of about 50 $\mu$ m. The pumping energy was about 1 nJ/pulse, and then the pump power density estimated from the spot size was about 13 $\mu$J/cm$^2$. By changing the optical path length of the probe beam, the reflectivity change $\Delta R/R$ was recorded as a function of the delay time. 

The THz-rate pulse trains were generated by Fourier transform (FT) pulse shaping techniques. This pulse shaping was performed by use of liquid-crystal spatial light modulators (LC-SLM), in which the applied electric field controls the index of refraction of each pixel and thus provides control over phase delay across the spectrum.\cite{rf14,rf15} The optical pulse shaper consists of a pair of 1200 line/mm gratings, a pair of 15-cm focal-length lenses, and LC-SLM.\cite{rf14} When using a single LC-SLM as a phase mask, it is difficult to obtain flat-topped shape. We used two LC-SLM$'$s to manipulate both the spectral phase and amplitude of the optical frequency components in the Fourier transformed plane.\cite{rf16,rf17} Combination of two SLM's enabled us to provide arbitrarily shaped temporal waveforms. Referring to the Fourier transform spectra of the prespecified temporal intensity profiles, we decided a mask pattern needed for the flat-topped shape. Almost flat-topped shape pulse trains were obtained by fine tuning of the optical layout of pulse shaper. The flat-topped pulse train as the excitation pulses enabled us to investigate whether the amplitude of coherent phonon increases linearly or nonlinearly when the pulse sequence is increased. The intensity profile of the shaped pulse was measured by a cross correlation method using second harmonic generation (SHG). We have generated the pulse trains at a repetition rate (a time interval of the pulse components) of 3.02 THz, 3.67 THz, and 4.40 THz which were matched with the $A_{1g}$-mode frequencies of Bi-Bi, Bi-Sb, and Sb-Sb vibrations, respectively. The pulse sequences from single to quintuple components were prepared for the pulse trains. 

\begin{figure}
\includegraphics[width=8.6cm]{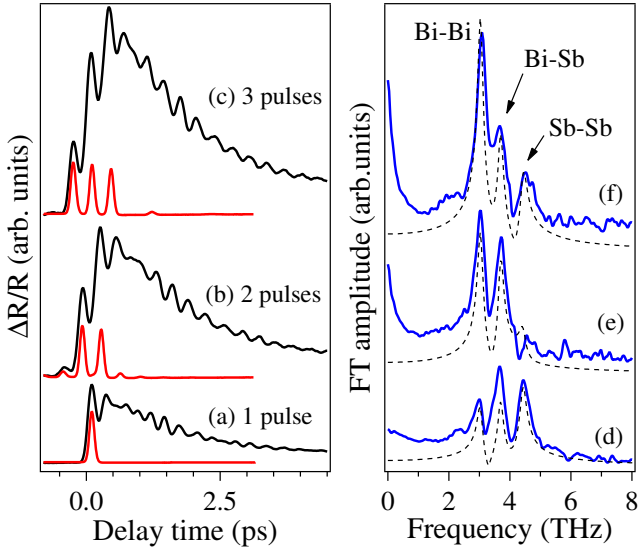}
\caption{Selective enhancement of the coherent $A_{1g}$ phonon of Bi-Bi vibrations in Bi$_{0.31}$Sb$_{0.69}$ mixed crystals at room temperature. (a)-(c) corresponding to the time domain data when the pulse sequence is varied from single to triple components. Dotted line represent SHG profile of the pulse sequence used. (d)-(f) represent FT spectra, and dotted line represent calculated spectra by using Eq. (1). 
}
\label{Fig1}
\end{figure}
Figure 1 demonstrates the result of the selective enhancement of the coherent Bi-Bi phonon by use of THz-rate pulse trains for different numbers of the pulse component. In Fig.1(a)-(c) the dotted line show SHG intensity profiles for the pulse train with a repetition rate of 3.02 THz ($T \sim$ 331 fs). The envelope of the pulse trains has almost flat-toped shape, and the repetition rates are well matched with the frequencies of the $A_{1g}$-phonon mode of Bi-Bi lattice vibration. The pulse width of each pulse component in the pulse train is about 70 fs. We also measured SHG signal for the THz-rate pulse trains at the repetition rate of 3.67 THz and 4.40 THz, and the profiles of the pulse trains showed also the flat-toped shape. In Fig.1 (a), $\Delta R/R$ induced by a single excitation pulse is shown. The oscillations due to the modulation of the dielectric function by coherent lattice vibrations appear on the background component that arises from photo-excited carriers. Figure 1 (d) shows the FT spectrum of the time domain data in Fig.1 (a). In this spectrum three peaks can be observed. They are assigned to Bi-Bi (3.02 THz), Bi-Sb (3.67 THz), and Sb-Sb (4.40 THz) phonons. These peak frequencies agree well with the results of Raman measurements for Bi-Sb mixed crystals.\cite{rf12} Figure 1 (b) and (c) show the reflectivity change induced by double pulses and triple pulses, respectively. The FT spectra of the signals in Fig.1 (b) and (c) are shown in Fig.1 (e) and (f), respectively. As the number of pulses increases, only the Bi-Bi phonon is enhanced, whereas the Sb-Sb phonon remains weak. If we assume that each mode observed in the Bi-Sb mixed crystal oscillates independently, the synthesized phonon by the pulse train is described by a linear combination of damped harmonic oscillations $R(t)$,\cite{rf9} 
\begin{equation}
R(t) = \sum_{m,n} A_{n}e^{-\gamma_{n}(t-m \Delta t)}\cos[\omega_{n} (t-m\Delta t)].
\end{equation}
Here, $A_{n}$ is the amplitude, $\gamma_{n}$ the decay rate, $\omega_{n}$ the frequency of $n$th mode, and $\Delta t$ is the time difference of the pulse sequence. The parameters $A_{n}$, $\gamma_{n}$, and $\omega_{n}$ are determined from fitting of the time domain wave in Fig.1(a) with a linear combination of the damped harmonic oscillations, although there is a slight discrepancy between the experimental data and fitting data. For Bi-Bi, Bi-Sb, and Sb-Sb phonons the amplitude ratios and the decay times are 1.0 : 3.0 : 2.8 and 1.96 ps, 0.95 ps, and 0.89 ps, respectively. We calculate the FT spectra in Fig.1(d)-(f) by using Eq.(1) and these fitting parameters. The calculated spectra reproduce the experimental data qualitatively, but there are slight differences in their ratios of FT amplitude and line width. These differences would result from some errors in the determination of the fitting parameters in the time domain. The experimental results for the selective enhancement by 3.02 THz pulse trains can be explained as follows. The amplitude of two coherent phonons are enhanced (canceled) when the two phonons are in phase (out of phase).\cite{rf9} There is a relationship between the period of the Bi-Bi and Sb-Sb vibrations, i.e., $T_{Bi} \approx 1.5 T_{Sb}$, where $T_{Bi}$ = 331 fs is the period of the $A_{1g}$ mode for the Bi-Bi vibration and $T_{Sb}$ = 227 fs is that for the Sb-Sb vibration. The above relations between the phonon periods and the repetition rate of the pulse train predict that the Sb-Sb vibrations generated by the double pulses can be canceled, while the Bi-Bi vibration is enhanced. When using the triple pulse, the Sb-Sb vibration would be generated again by the third pulse component. On the other hands, the ratio of the periods of the Bi-Sb vibration and the other vibrations is not described by simple integer numbers. When using the 3.02 THz pulse train, the Bi-Sb vibrations generated by multiple pulse components are not in phase and then are not enhanced. The same results for selective enhancement of the Bi-Sb and Sb-Sb vibrations are obtained.

\begin{figure}
\includegraphics[width=8.6cm]{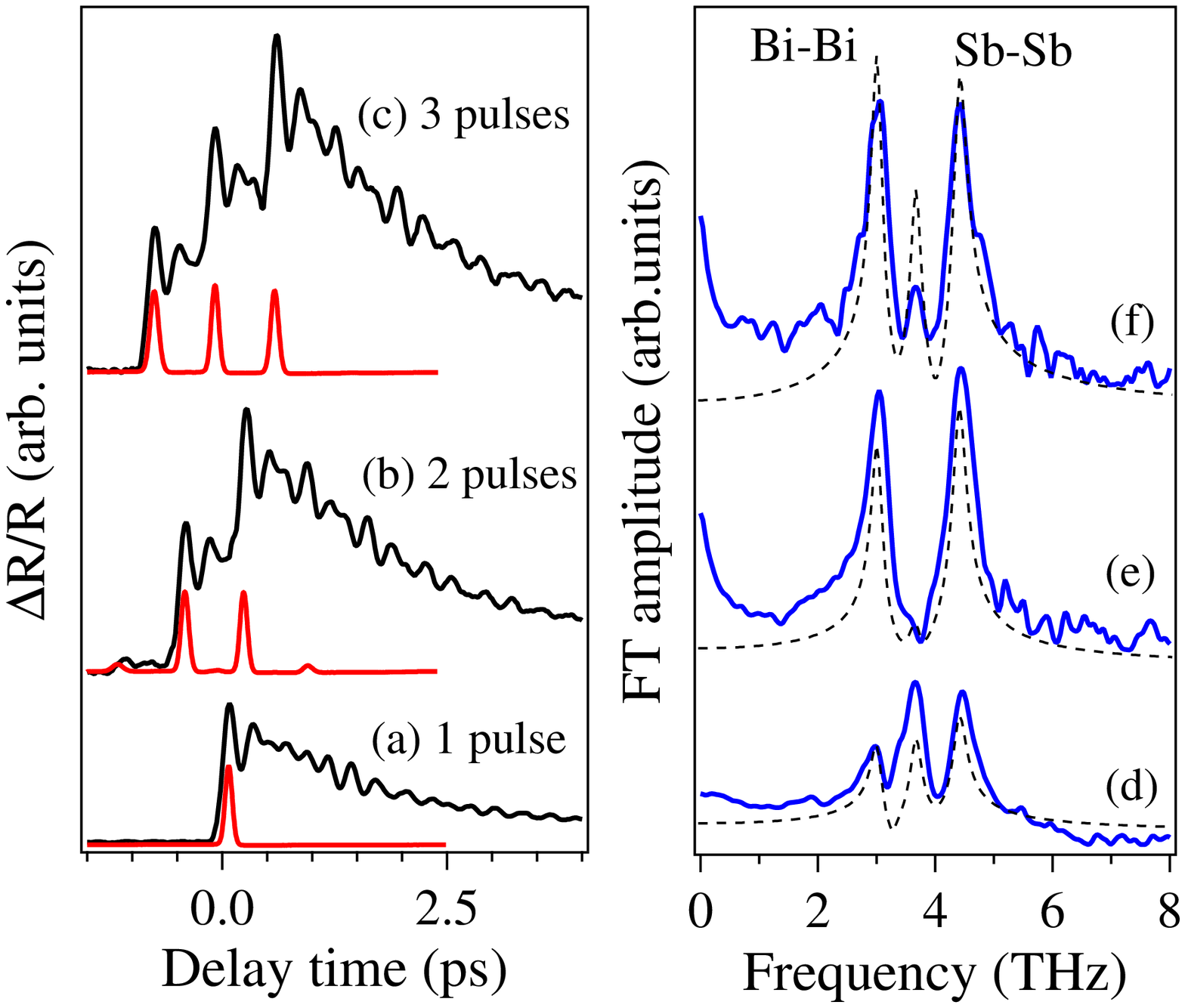}
\caption{Selective cancellation of the coherent $A_{1g}$ phonon of Bi-Sb vibrations in Bi$_{0.31}$Sb$_{0.69}$ mixed crystals at room temperature. (a)-(c) corresponding to the time domain data when the pulse sequence is varied from single to triple components. Doted line represent SHG profile of the pulse sequence used. (d)-(f) represent FT spectra, and dotted line represent calculated spectra by using Eq. (1).
}
\label{Fig2}
\end{figure}
Selective cancellation of the coherent Bi-Sb vibration are performed using pulse trains at the repetition rate of 6.04 THz ($T_{p}$ = 662 fs) which satisfies the relation $T_{p} = 2T_{Bi} \approx 3T_{Sb}$. Figure 2 demonstrates the result of the selective cancellation of the coherent Bi-Sb vibration when the number of pulse sequences is varied. In Fig.2 (a)-(c), the reflectivity changes $\Delta R/R$ are shown as a function of delay time, where the dotted line shows SHG intensity profiles for the pulse train at a repetition rate of 6.04 THz ($T \sim$ 662\,fs). The FT spectra of Fig.2 (a)-(c) are shown in Fig.2 (d)-(f), respectively. As shown in Fig.2 (e) when the pulse sequence is double pulses, only the Bi-Sb vibration is clearly canceled. However, as shown in Fig.2 (f), discrepancy between the experimental data and fitting profiles is large for the Bi-Sb vibration (4.40 THz) when the pulse sequence is triple. This discrepancy would arise mainly from the ambiguity in the time-domain fitting in Fig.1(a). The FT spectra calculated for a linear combination of damped harmonic oscillators, which are also shown in Fig.2 (d)-(f), well reproduce the experimental data except for Fig.2 (f). These results suggest that a precise manipulation of phonons excited coherently is possible.

\begin{figure}
\includegraphics[width=8.6cm]{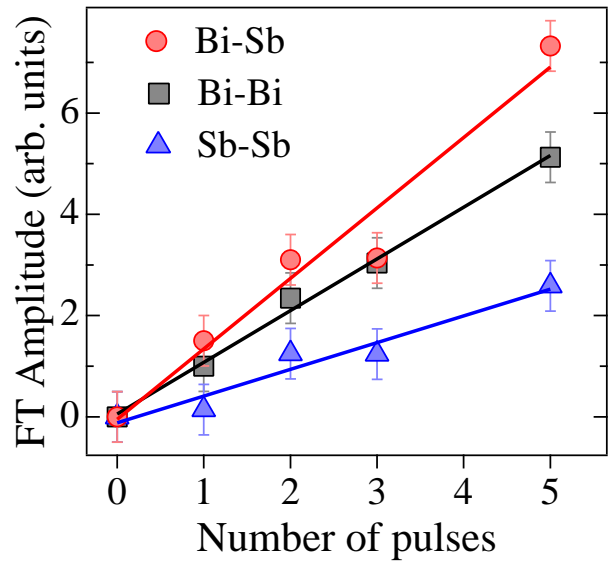}
\caption{The relation between the number of pulse sequence and the FT amplitude of coherent phonons.
}
\label{Fig3}
\end{figure}
In order to estimate the ratios of the enhanced amplitude to the number of pulse sequence, we have normalized the amplitude with the peak intensity of the flat-toped pulse sequences. Figure 3 shows the relation between the number of pulse sequence in the THz-rate pulse train and the FT amplitude of the coherent phonons. The FT amplitudes increase linearly as the number of the pulse sequence increases. In our experiment, with pumping energy below 13 $\mu$J/cm$^2$, saturation of the amplitude of coherent phonons was not observed. 

In conclusion we have demonstrated both selective enhancement and cancellation of coherent optical phonons by use of THz-rate pulse trains in Bi-Sb mixed crystal systems. The Bi-Bi, Bi-Sb, and Sb-Sb vibrations can be selectively enhanced and canceled for pulse trains with an appropriate repetition rate. The amplitudes of the $A_{1g}$ coherent phonons, which are excited by the pulse train, grow linearly as the number of pulse sequence increases. Our results suggest that the optical manipulation of lattice vibrations would be realized in ultrafast time scale.
 
We would like to thank Kansai Advanced Research Center, CRL, where first stage of this work was done. This work was partially supported by a Grant-in-Aid for the Scientific Research from the Ministry of Education, Science, Sports, and Culture of Japan. The experiments were carried out in the Venture Business Laboratory, Osaka University.

\end{document}